\documentclass{article}
\usepackage{amsmath,amssymb,
bm,graphicx}

\def\abstract#1{\vskip 7mm
        \begin{center}{\large Abstract}\par \smallskip
                \begin{minipage}[c]{12cm}
                        \small #1
                \end{minipage}
        \end{center}
}
\def\title#1{\begin{center}{\Large\bf #1}\end{center}}
\def\author#1{\vskip 5mm \begin{center}{#1}\end{center}}
\def\address#1{\begin{center}{\it #1}\end{center}}
\makeatletter
\@ifundefined{lesssim}{}{}
\@ifundefined{gtrsim}{}{}
\def\vereq#1#2{\lower3pt\vbox{\baselineskip1.5pt \lineskip1.5pt
\ialign{$\m@th#1\hfill##\hfil$\crcr#2\crcr\sim\crcr}}}
\makeatother

\begin{document}

\def\tr{{\rm tr}\,}

\def\be{\begin{equation}}
\def\ee{\end{equation}}
\def\bea{\begin{eqnarray}}
\def\eea{\end{eqnarray}}
\def\beaa{\begin{eqnarray*}}
\def\eeaa{\end{eqnarray*}}
\def\nn{\nonumber \\}
\def\cF{{\cal F}}
\def\det{{\rm det\,}}
\def\Tr{{\rm Tr\,}}
\def\e{{\rm e}}

\title{%
 Dark energy, inflation and dark matter from modified $F(R)$ gravity
}
\author{%
  Shin'ichi Nojiri\footnote{E-mail:nojiri@phys.nagoya-u.ac.jp}
  and
  Sergei D. Odintsov\footnote{E-mail:odintsov@ieec.uab.es, also at Lab. Fundam. Study, Tomsk State
Pedagogical University, Tomsk}
}
\address{%
  $^1$Department of Physics, Nagoya University,
   Nagoya 464--8602, Japan\\
  $^2$Instituci\`{o} Catalana de Recerca i Estudis Avan\c{c}ats (ICREA)
and Institut de Ciencies de l'Espai (IEEC-CSIC),
Campus UAB, Facultat de Ciencies, Torre C5-Par-2a pl, E-08193 Bellaterra
(Barcelona), Spain
}

\abstract{
We review modified $F(R)$ gravity as realistic candidate to describe the
observable universe expansion history.
 We show that  recent cosmic acceleration,
radiation/matter-dominated epoch and  inflation could be realized in
the framework of $F(R)$-gravity in the unified way.
For some viable classes of $F(R)$-gravity,  the Newton
law
is respected and there is no so-called matter instability (the very
heavy positive mass for additional scalar degree of freedom is generated).
The reconstruction program in modified gravity is also reviewed and it is
demonstrated that {\it any} time-evolution
of the universe expansion  could be realized in
$F(R)$-gravity. These models remain to be realistic also in the presence
of
non-minimal gravitational coupling with usual matter. It is shown that
same model which passes local tests and predicts the unification of
inflation with cosmic acceleration also describes dark matter thanks to
presence of additional scalar degree of freedom and chameleon mechanism.}

\section{Introduction \label{Sec1}}

Modified gravity suggests very natural answers to resolution of several
fundamental cosmological problems. For instance, the observable universe
expansion history may be described by modified gravity. Indeed, it gives
very beautiful unification of the early-time inflation and late-time
acceleration thanks to different role of gravitational terms relevant at
small and at large curvature. Moreover, the coincidence problem may be
solved in such theory simply by the universe expansion. Some models of
modified gravity are predicted by string/M-theory considerations.

 From another side, dark matter may be described totally in terms of
modified gravity. Moreover, modified gravity may be useful in high energy
physics, for instance, to solve the hierarchy or gravity-GUTs unification
problems. Finally, modified gravity may pass the local tests and
cosmological bounds.

Usually the evolution of the universe can be described by the FRW equation:
\be
\label{JGRG1}
\frac{3}{\kappa^2}H^2 = \rho\ .
\ee
Here  the spatial part of the universe is assumed to be flat.
We denote the Hubble rate by $H$, which is defined in terms of the scale factor $a$ by
\be
\label{JGRG2}
H\equiv \frac{\dot a}{a}\ .
\ee
In (\ref{JGRG1}), $\rho$ expresses the energy density of the usual matter,
dark matter,
and dark energy.
The dark energy could be cosmological constant and/or a matter with `equation of state
(EoS)' parameter $w$, which is less than $-1/3$ and is defined by
\be
\label{JGRG3}
w\equiv \frac{p}{\rho} \ .
\ee
Instead of including unknown exotic matter or energy, one may consider the
modification of gravity,
which corresponds to the change of the l.h.s. in (\ref{JGRG1}).

An example of such modified gravity pretending to describe dark energy
could be the
scalar-Einstein-Gauss-Bonnet gravity \cite{GB}, whose
action is given by
\be
\label{JGRG4}
S=\int d^4x \sqrt{-g}\left\{ \frac{1}{2\kappa^2}R
 - \frac{1}{2}\partial_\mu \phi \partial^\mu \phi
 - V(\phi) + f(\phi) {\cal G}\right\} \ .
\ee
Here ${\cal G}$ is  Gauss-Bonnet invariant:
\be
\label{JGRG5}
{\cal G}\equiv R^2 - 4 R_{\mu\nu} R^{\mu\nu} + R_{\mu\nu\rho\sigma}R^{\mu\nu\rho\sigma}\ .
\ee

Another example is so-called $F(R)$-gravity (for a review, see \cite{review}).
In  $F(R)$-gravity models\cite{review}, the scalar curvature $R$ in
the Einstein-Hilbert action
\be
\label{JGRG6}
S_{\rm EH}=\int d^4 x \sqrt{-g} R\ ,
\ee
is replaced by a proper function of the scalar curvature:
\be
\label{JGRG7}
S_{F(R)}=\int d^4 x \sqrt{-g} F(R)\ .
\ee

Recently, an interesting realistic theory has been proposed in \cite{HS},
where $F(R)$ is given by
\be
\label{JGRG8}
F(R)=\frac{1}{2\kappa^2}\left( R + f_{HS}(R) \right)\ ,\quad
f_{HS}(R)=-\frac{m^2 c_1 \left(R/m^2\right)^n}{c_2 \left(R/m^2\right)^n + 1}\ .
\ee
In this model, $R$ is large even in the present universe, and $f_{HS}(R)$
could be expanded
by the inverse power series of $R$:
\be
\label{JGRG9}
f_{HS}(R)\sim - \frac{m^2 c_1}{c_2} + \frac{m^2 c_1}{c_2^2}
\left(\frac{R}{m^2}\right)^{-n} + \cdots \ ,
\ee
Then there appears an effective cosmological constant $\Lambda_{\rm eff}$ as
$\Lambda_{\rm eff} = m^2 c_1/c_2$, which generates the accelerating expansion
in the present universe

In the HS-model (\ref{JGRG8}), there occurs a flat spacetime solution,
where $R=0$, since the
following condition is satisfied:
\be
\label{JGRG10}
\lim_{R\to 0} f_{HS}(R) = 0\ .
\ee
An interesting point in the HS model is that several cosmological
conditions could be satisfied.

In the next section, we review on the general properties of  $F(R)$-gravity.
After some versions of $F(R)$-gravity were proposed as a model of the dark
energy, there were indicated several problems/viability criteria,
which we review in Section \ref{Sec3}. It is shown how the critique of
modified gravity may be removed for realistic models.
In Section \ref{Sec4}, we propose models  \cite{Nojiri:2007as} and \cite{Nojiri:2007cq},
which unify the early-time inflation
and the recent cosmic acceleration and
 pass several cosmological constraints.
Reconstruction program for  $F(R)$-gravity  is reviewed in Section
\ref{Sec5}. The partial reconstruction scenario is proposed.
Section six is devoted to the description of dark matter in terms of
viable
modified gravity where composite scalar particle from $F(R)$ gravity plays
the role of dark particle. Non-minimal modified gravity is discussed in
section seven. Some summary and outlook is given in the last
section.

\section{General properties of $F(R)$-gravity \label{Sec2}}

In this section, the general properties of the $F(R)$-gravity are reviewed.
For general $F(R)$-gravity, one can define an effective equation of state
(EoS) parameter.
The FRW equations in Einstein gravity coupled with perfect fluid are:
\be
\label{JGRG11}
\rho=\frac{3}{\kappa^2}H^2 \ ,\quad p= - \frac{1}{\kappa^2}\left(3H^2 + 2\dot H\right)\ .
\ee
For modified gravities, one may define an effective EoS
parameter as follows:
\be
\label{JGRG12}
w_{\rm eff}= - 1 - \frac{2\dot H}{3H^2} \ .
\ee

The equation of motion for modified gravity
is given by
\be
\label{JGRG13}
\frac{1}{2}g_{\mu\nu} F(R) - R_{\mu\nu} F'(R) - g_{\mu\nu} \Box F'(R)
+ \nabla_\mu \nabla_\nu F'(R) = - \frac{\kappa^2}{2}T_{(m)\mu\nu}\ .
\ee
By assuming spatially flat FRW universe,
\be
\label{JGRG14}
ds^2 = - dt^2 + a(t)^2 \sum_{i=1,2,3} \left(dx^i\right)^2\ ,
\ee
 the FRW-like equation follows:
\be
\label{JGRG15}
0 = - \frac{F(R)}{2} + 3\left(H^2 + \dot H \right) F'(R) - 18 \left(4H^2 \dot H
+ H \ddot H \right) F''(R)
+ \kappa^2 \rho_{(m)}
\ee

There may be several (often exact) solutions of (\ref{JGRG15}).
Without any matter,  assuming that the Ricci tensor could be covariantly
constant,
that is,  $R_{\mu\nu}\propto g_{\mu\nu}$, Eq.(\ref{JGRG13}) reduces to the
algebraic equation:
\be
\label{JGRG16}
0 = F(R) - 2 R F(R)\ .
\ee
If Eq.(\ref{JGRG16}) has a solution, the Schwarzschild (or Kerr) - (anti-)de Sitter space is
an exact vacuum solution (see\cite{eli} and refs. therein).

When $F(R)$ behaves as $F(R) \propto R^m$ and there is no matter, there appears the following solution:
\be
\label{JGRG17}
H \sim \frac{-\frac{(m-1)(2m-1)}{m-2}}{t}\ ,
\ee
which gives the following effective EoS parameter:
\be
\label{JGRG18}
w_{\rm eff}=-\frac{6m^2 - 7m - 1}{3(m-1)(2m -1)}\ .
\ee

When $F(R) \propto R^m$ again but if the matter with a constant EoS
parameter $w$ is included,
one may get the following solution:
\be
\label{JGRG19}
H \sim \frac{\frac{2m}{3(w+1)}}{t}\ ,
\ee
and the effective EoS parameter is given by
\be
\label{JGRG20}
w_{\rm eff}= -1 + \frac{w+1}{m}\ .
\ee
This shows that modified gravity may describe early/late-time universe
acceleration.

\section{Problems with  $F(R)$-gravity \label{Sec3}}

Immediately, after the $F(R)$-models were proposed as models of the dark
energy,
 there appeared several works
 \cite{Chiba,DK} (and more recently in \cite{Kami,Chiba2}) criticizing
such theories.

First of all, we comment on the claim in \cite{Chiba}.
Note that one can rewrite $F(R)$-gravity in the scalar-tensor form.
By introducing the auxiliary field $A$, we rewrite the action (\ref{JGRG7}) of the $F(R)$-gravity
in the following form:
\be
\label{JGRG21}
S=\frac{1}{\kappa^2}\int d^4 x \sqrt{-g} \left\{F'(A)\left(R-A\right) + F(A)\right\}\ .
\ee
By the variation over $A$, one obtains $A=R$. Substituting $A=R$ into
the action (\ref{JGRG21}),
one can reproduce the action in (\ref{JGRG7}). Furthermore, we rescale the
metric in the following way (conformal transformation):
\be
\label{JGRG22}
g_{\mu\nu}\to \e^\sigma g_{\mu\nu}\ ,\quad \sigma = -\ln F'(A)\ .
\ee
Hence, the Einstein frame action is obtained:
\bea
\label{JGRG23}
S_E &=& \frac{1}{\kappa^2}\int d^4 x \sqrt{-g} \left( R - \frac{3}{2}g^{\rho\sigma}
\partial_\rho \sigma \partial_\sigma \sigma - V(\sigma)\right) \ ,\nn
V(\sigma) &=& \e^\sigma g\left(\e^{-\sigma}\right)
 - \e^{2\sigma} f\left(g\left(\e^{-\sigma}\right)\right) = \frac{A}{F'(A)} - \frac{F(A)}{F'(A)^2}
\eea
Here $g\left(\e^{-\sigma}\right)$ is given by solving the equation
$\sigma = -\ln\left( 1 + f'(A)\right)=\ln F'(A)$ as $A=g\left(\e^{-\sigma}\right)$.
Due to the scale transformation (\ref{JGRG22}), there appears a coupling
of the scalar
field $\sigma$ with usual matter.
The mass of $\sigma$ is given by
\be
\label{JGRG24}
m_\sigma^2 \equiv \frac{1}{2}\frac{d^2 V(\sigma)}{d\sigma^2}
=\frac{1}{2}\left\{\frac{A}{F'(A)} - \frac{4F(A)}{\left(F'(A)\right)^2} + \frac{1}{F''(A)}\right\}\ .
\ee
Unless $m_\sigma$ is very  large, there  appears the large
correction to the Newton law.
Naively, one expects the order of the mass $m_\sigma$ could be that of the
Hubble rate, that is,
$m_\sigma \sim H \sim 10^{-33}\,{\rm eV}$, which is very light and the correction could be
very large, which is the claim in \cite{Chiba}.

We should note, however, that the mass $m_\sigma$  depends
on the detailed form of $F(R)$
in general \cite{NO}.
Moreover, the mass $m_\sigma$  depends on the curvature.
The curvature on the earth $R_{\rm earth}$ is much larger than the average curvature $R_{\rm solar}$
in the solar system and $R_{\rm solar}$ is also much larger than the average curvature in
the unverse, whose order is given by the square of the Hubble rate $H^2$, that is,
$R_{\rm earth} \gg R_{\rm solar} \gg H^2$.
Then if the mass becomes large when the curvature is large, the correction to the Newton
law could be small.
Such a mechanism is called the Chameleon mechanism and proposed for the
scalar-tensor theory
in \cite{KW}. In fact, the HS model \cite{HS} has this property
 and the correction to the Newton
law can be very small on the earth or in the solar system.
In the HS model,  the mass $m_\sigma$ is given by (see also
\cite{newton})
\be
\label{JGRG25}
m_\sigma^2 \sim \frac{m^2 c_2^2}{2 n(n+1) c_1}\left(\frac{R}{m^2}\right)^{n+2}\ .
\ee
Here the order of the mass-dimensional parameter $m^2$ could be $m^2\sim 10^{-64}\,{\rm eV}^2$.
Then in solar system, where $R\sim 10^{-61}\,{\rm eV}^2$, the mass is given by
$m_\sigma^2 \sim 10^{-58 + 3n}\,{\rm eV}^2$ and in the air on the earth,
where $R \sim 10^{-50}\,{\rm eV}^2$, $m_\sigma^2 \sim 10^{-36 + 14n}\,{\rm eV}^2$.
The order of the radius of the earth is $10^7\,{\rm m} \sim \left(10^{-14}\,{\rm eV}\right)^{-1}$.
Therefore the scalar field $\sigma$ could be heavy enough if $n\gg 1$ and the correction to the
Newton law is not observed being extremely small.
On the other hand, in the air on the earth, if we choose $n=10$, for example,
 one gets the mass
is extremely large:
\be
\label{JGRG26}
m_\sigma \sim 10^{43}\,{\rm GeV} \sim 10^{29} \times M_{\rm Planck}\ .
\ee
Here $M_{\rm Planck}$ is the Planck mass. Hence, the Newton law correction
should be extremely small.

Let us discuss the matter instability proposed in \cite{DK}, which may
appear when the
energy density or the curvature is large compared with the average one in
the universe,
as is the case inside of the planet.
Multiplying $g^{\mu\nu}$ with Eq.(\ref{JGRG13}), one obtains
\be
\label{JGRG27}
\Box R + \frac{F^{(3)}(R)}{F^{(2)}(R)}\nabla_\rho R \nabla^\rho R
+ \frac{F'(R) R}{3F^{(2)}(R)} - \frac{2F(R)}{3 F^{(2)}(R)}
= \frac{\kappa^2}{6F^{(2)}(R)}T\ .
\ee
Here $T$ is the trace of the matter energy-momentum tensor:
$T\equiv T_{(m)\rho}^{\ \rho}$. We also denote $d^nF(R)/dR^n$ by $F^{(n)}(R)$.
Let us now consider the perturbation from the solution of the Einstein
gravity.
We denote the scalar curvature solution given by the matter density in the
Einstein gravity by
$R_b\sim (\kappa^2/2)\rho>0$ and separate the scalar curvature $R$ into the sum of $R_b$
and the perturbed part $R_p$ as $R=R_b + R_p$ $\left(\left|R_p\right|\ll \left|R_b\right|\right)$.
Then Eq.(\ref{JGRG27}) leads to the perturbed equation:
\bea
\label{JGRG28}
0 &=& \Box R_b + \frac{F^{(3)}(R_b)}{F^{(2)}(R_b)}\nabla_\rho R_b \nabla^\rho R_b
+ \frac{F'(R_b) R_b}{3F^{(2)}(R_b)} \nn
&& - \frac{2F(R_b)}{3 F^{(2)}(R_b)} - \frac{R_b}{3F^{(2)}(R_b)} + \Box R_p
+ 2\frac{F^{(3)}(R_b)}{F^{(2)}(R_b)}\nabla_\rho R_b \nabla^\rho R_p + U(R_b) R_p\ .
\eea
Here $U(R_b)$ is given by
\bea
\label{JGRG29}
U(R_b) &\equiv& \left(\frac{F^{(4)}(R_b)}{F^{(2)}(R_b)} - \frac{F^{(3)}(R_b)^2}{F^{(2)}(R_b)^2}\right)
\nabla_\rho R_b \nabla^\rho R_b + \frac{R_b}{3} \nn
&& - \frac{F^{(1)}(R_b) F^{(3)}(R_b) R_b}{3 F^{(2)}(R_b)^2} - \frac{F^{(1)}(R_b)}{3F^{(2)}(R_b)}
+ \frac{2 F(R_b) F^{(3)}(R_b)}{3 F^{(2)}(R_b)^2} - \frac{F^{(3)}(R_b) R_b}{3 F^{(2)}(R_b)^2}
\eea
It is convenient to consider the case that $R_b$ and $R_p$ are uniform,
that is, they do not depend on the
spatial coordinate. Hence, the d'Alembertian can be replaced with the
second derivative with respect
to the time coordinate: $\Box R_p \to - \partial_t^2 R_p$ and Eq.(\ref{JGRG29}) has
the following structure:
\be
\label{JGRG30}
0=-\partial_t^2 R_p + U(R_b) R_p + {\rm const.}\ .
\ee
Then if $U(R_b)>0$, $R_p$ becomes exponentially large with time $t$:
$R_p\sim \e^{\sqrt{U(R_b)} t}$ and the system becomes unstable.
In the $1/R$-model \cite{CDTT}, since the order of mass parameter $m_\mu$ is
\be
\label{JGRG31}
\mu^{-1}\sim 10^{18} \mbox{sec} \sim \left( 10^{-33} \mbox{eV} \right)^{-1}\ ,
\ee
one finds
\bea
\label{JGRG32}
&& U(R_b) = - R_b + \frac{R_b^3}{6\mu^4} \sim \frac{R_0^3}{\mu^4}
\sim \left(10^{-26} \mbox{sec}\right)^{-2} \left(\frac{\rho_m}{\mbox{g\,cm}^{-3}}\right)^3\ ,\nn
&& R_b \sim  \left(10^3 \mbox{sec}\right)^{-2} \left(\frac{\rho_m}{\mbox{g\,cm}^{-3}}\right)
\eea
Hence, the model is unstable and it would decay in $10^{-26}$ sec (for
planet size).
On the other hand, in $1/R + R^2$-model \cite{NO}, we find
\be
\label{JGRG33}
U(R_0)\sim \frac{R_0}{3}>0\ .
\ee
Then the system could be unstable again but the decay time is $\sim$
$1,000$ sec, that is,
macroscopic.
In HS model \cite{HS}, $U(R_b)$ is negative\cite{newton}:
\be
\label{JGRG34}
U(R_0) \sim - \frac{(n+2)m^2 c_2^2}{c_1 n(n+1)} < 0 \ .
\ee
Therefore, there is no matter instability\cite{newton}.

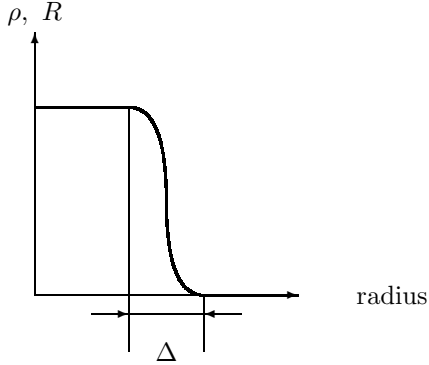
\begin{figure}

\begin{center}

\unitlength=0.5mm
\begin{picture}(120,100)

\thinlines

\put(10,20){\vector(1,0){70}}
\put(10,20){\vector(0,1){70}}

\put(35,70){\line(0,-1){65}}
\put(55,20){\line(0,-1){15}}
\put(25,15){\vector(1,0){10}}
\put(65,15){\vector(-1,0){10}}
\put(35,15){\line(1,0){20}}

\thicklines

\put(10,95){\makebox(0,0){$\rho,\ R$}}
\put(105,20){\makebox(0,0){radius}}

\put(10,70){\line(1,0){25}}
\put(55,20){\line(1,0){25}}
\qbezier(35,70)(45,70)(45,45)
\qbezier(45,45)(45,20)(55,20)

\put(45,5){\makebox(0,0){$\Delta$}}

\end{picture}

\end{center}

\caption{\label{fig1} Typical behavior of $R$ and $\rho$ near the surface of the stellar configuration.}
\end{figure}

Let us discuss the critical claim against modified gravity in
\cite{Kami,Chiba2}.
As shown in (\ref{JGRG16}), as an exact solution, there appears de Sitter-Schwarzschild spacetime
in $F(R)$-gravity. The claim in \cite{Kami,Chiba2} is that the solution does not match onto
the stellar interior solution.
Since it is difficult to construct explicit solution describing the stellar configuration even
in the Einstein gravity, we now proceed in the following way:
First, we separate $F(R)$ into the sum of the Einstein-Hilbert part and
other part as $F(R)=R+f(R)$.
Then Eq.(\ref{JGRG13}) has the following form:
\bea
\label{JGRG35}
&& \frac{1}{2}g_{\mu\nu} R - R_{\mu\nu} - \frac{1}{2} g_{\mu\nu} \Lambda
+ \frac{\kappa^2}{2}T_{(m)\mu\nu} \nn
&& = - \frac{1}{2}g_{\mu\nu} \left( f(R) + \Lambda \right) + R_{\mu\nu} f'(R)
+ g_{\mu\nu}\Box f'(R) - \nabla_\mu \nabla_\nu f'(R)\ .
\eea
Here $-\Lambda$ is the value of $f(R)$ in the present universe, that is, $\Lambda$ is the
effective cosmological constant: $\Lambda = - f(R_0)$.
We now treat the r.h.s. in (\ref{JGRG35}) as a perturbation.
Then the last two derivative terms in (\ref{JGRG35}) could be dangerous
since there could be jump in the value of the scalar curvature $R$ on the surface of stellar
configuration.
Of course, the density on the surface could change in a finite width
$\Delta$ as in Figure \ref{fig1}
and the derivatives should be finite and the magnitude could be given by
\be
\label{JGRG36}
\partial_\mu \sim \frac{1}{\Delta}\ .
\ee

One now assumes the order of the derivative could be the order of the
Compton length of proton:
\be
\label{JGRG37}
\partial_\mu \sim m_p \sim 1\,{\rm GeV}\sim 10^9\,{\rm eV}
\ee
Here $m_p$ is the mass of proton. It is also assumed
\be
\label{JGRG38}
R\sim R_e \sim 10^{-47}\,{\rm eV}^2\ ,
\ee
that is, the order of the scalar curvature $R$ is given by the order of it inside the earth.

In case of the $1/R$ model \cite{CDTT}, one gets
\be
\label{JGRG39}
\Box f'(R) \sim \nabla_\mu \nabla_\nu f'(R) \sim \frac{m_p^2 \mu^4}{R^2}
\sim 10^{-20}\,{\rm eV}^2\gg R_e\ .
\ee
Then the perturbative part could be much larger than unperturbative part in (\ref{JGRG35}),
say, $R\sim R_e \sim 10^{-47}\,{\rm eV}^2$.
Therefore, the perturbative expansion could be inconsistent.

In case of the model \cite{HS}, however, we find
\be
\label{JGRG40}
\Box f'(R) \sim \nabla_\mu \nabla_\nu f'(R) \sim
\frac{m_p^2 \Lambda}{m^2}  \left(\frac{R}{m^2}\right)^{-n-1} \sim 10^{-3 - 17n}\,{\rm eV}^2\ .
\ee
Then if $n>2$, we find $\Box f'(R)$, $\nabla_\mu \nabla_\nu f'(R)\ll R_e$ and therefore
the perturbative expansion could be consistent. This indicates that such
modified gravity model may pass the above test. Thus, it is demonstrated
that some versions of modified gravity may easily pass above tests.

\section{Unifying inflation and cosmic acceleration \label{Sec4}}

In this section, we consider an extension of the HS model \cite{HS}
 to unify the early-time inflation
and late-time acceleration, following proposals
 \cite{Nojiri:2007as, Nojiri:2007cq}.

In order to construct such models, we impose the following conditions:
\begin{itemize}
\item Condition  that  inflation  occurs:
\be
\label{JGRG41}
\lim_{R\to\infty} f (R) = - \Lambda_i\ .
\ee
Here $\Lambda_i$ is an effective early-time cosmological constant.

Instead of (\ref{JGRG41}) one may impose the following condition
\be
\label{JGRG42}
\lim_{R\to\infty} f (R) = \alpha R^m\ .
\ee
Here $m$ and $\alpha$ are positive constants.
Then as shown in (\ref{JGRG19}), the scale factor $a(t)$  evolves as
\be
\label{JGRG43}
a(t) \propto t^{h_0}\ ,\quad h_0 \equiv \frac{2m}{3(w+1)}\ ,
\ee
and  $w_{\rm eff}= -1 + 2/3h_0$.
Here $w$ is the matter EoS parameter, which could correspond to dust or
radiation.
We assume $m\gg 1$ so that $\dot H/H^2 \gg 1$.
\item The condition that there is flat spacetime solution is given
as
\be
\label{JGRG44}
f(0)=0\
\ee
\item The condition that late-time acceleration occurs should be
\be
\label{JGRG45}
f(R_0)= - 2\tilde R_0\ ,\quad f'(R_0)\sim 0\ .
\ee
Here $R_0$ is the current curvature of the universe and we assume $R_0> \tilde R_0$.
Due to the condition (\ref{JGRG45}), $f(R)$ becomes almost constant in the present universe
and plays the role of the effective small cosmological constant:
$\Lambda_l \sim - f(R_0) = 2\tilde R_0$.
\end{itemize}

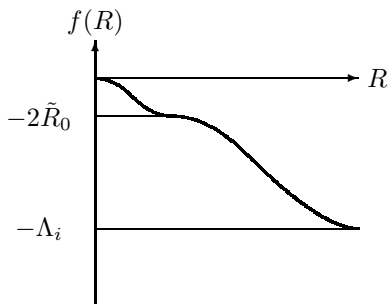
\begin{figure}

\begin{center}

\unitlength 0.5mm

\begin{picture}(100,100)

\put(10,70){\vector(1,0){70}}
\put(10,10){\vector(0,1){70}}

\put(10,85){\makebox(0,0){$f(R)$}}
\put(85,70){\makebox(0,0){$R$}}
\put(-5,60){\makebox(0,0){$-2\tilde R_0$}}
\put(-5,30){\makebox(0,0){$-\Lambda_i$}}

\thicklines

\qbezier(10,70)(15,70)(20,65)
\qbezier(20,65)(25,60)(30,60)
\qbezier(30,60)(40,60)(50,50)
\qbezier(50,50)(70,30)(80,30)

\thinlines

\put(10,60){\line(1,0){20}}
\put(10,30){\line(1,0){70}}

\end{picture}

\end{center}

\caption{\label{fig2} The typical behavior of $f(R)$ which satisfies the conditions (\ref{JGRG41}),
(\ref{JGRG44}), and (\ref{JGRG45}). }

\end{figure}

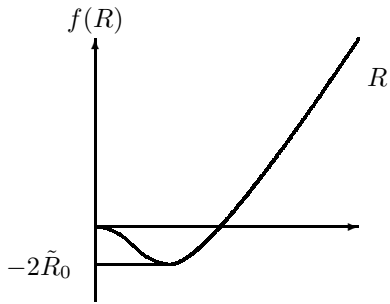
\begin{figure}

\begin{center}

\unitlength 0.5mm

\begin{picture}(100,100)

\put(10,30){\vector(1,0){70}}
\put(10,10){\vector(0,1){70}}

\put(10,85){\makebox(0,0){$f(R)$}}
\put(85,70){\makebox(0,0){$R$}}
\put(-5,20){\makebox(0,0){$-2\tilde R_0$}}

\thicklines

\qbezier(10,30)(15,30)(20,25)
\qbezier(20,25)(25,20)(30,20)
\qbezier(30,20)(40,20)(80,80)

\thinlines

\put(10,20){\line(1,0){20}}

\end{picture}

\end{center}

\caption{\label{fig3} The typical behavior of $f(R)$ which satisfies the conditions (\ref{JGRG42}),
(\ref{JGRG44}), and (\ref{JGRG45}). }

\end{figure}

The typical behavior of $f(R)$ which satisfies the conditions (\ref{JGRG41}),
(\ref{JGRG44}), and (\ref{JGRG45}) is given in Figure \ref{fig2} and the
behavior of $f(R)$
satisfying (\ref{JGRG42}), (\ref{JGRG44}), and (\ref{JGRG45}) is given in
Figure \ref{fig3}.

Some examples may be of interest.
An example which satisfies the conditions  (\ref{JGRG41}),
(\ref{JGRG44}), and (\ref{JGRG45}) is given by the following action\cite{Nojiri:2007as}:
\be
\label{JGRG46}
f(R) = - \frac{\left(R-R_0\right)^{2n+1} + R_0^{2n+1}}{f_0
+ f_1 \left\{\left(R-R_0\right)^{2n+1} + R_0^{2n+1}\right\}}\ .
\ee
Here $n$ is a positive integer.
The conditions (\ref{JGRG42}) and (\ref{JGRG45}) require
\be
\label{JGRG47}
\frac{R_0^{2n+1}}{f_0 +f_1  R_0^{2n+1}}=2\tilde R_0\ ,\quad
\frac{1}{f_1} = \Lambda_i\ .
\ee
One can now investigate how the exit from the inflation could be realized
in the model (\ref{JGRG46}).
It is easier to consider this problem in the scalar-tensor form (Einstein
frame) in (\ref{JGRG23}).
In the inflationary epoch, when the curvature $R=A$ is large, $f(R)$ has
the following form:
\be
\label{JGRG48}
f(R) \sim - \frac{1}{f_1} + \frac{f_0}{f_1^2 R^{2n+1}} \ .
\ee
Hence, one gets
\be
\label{JGRG49}
\sigma \sim \frac{(2n+1)f_0}{f_1^2 A^{2n+2}} \ ,
\ee
and
\be
\label{JGRG50}
V(\sigma) \sim \frac{1}{f_1} - \frac{2(n+1)f_0}{f_1^2}
\left(\frac{f_1^2 \sigma}{(2n+1)f_0}\right)^{\frac{2n+1}{2n+2}}\ .
\ee
Note that the scalar field $\sigma$ is dimensionless now.
Let us check the condition for the slow roll, $\left|V'/V\right|\ll 1$.
Since
\be
\label{JGRG51}
\frac{V'(\sigma)}{V(\sigma)} \sim - f_1 \left(\frac{f_1^2 \sigma}{(2n+1)f_0}\right)^{-\frac{1}{2n+2}}\ ,
\ee
if we start with $\sigma \sim 1$, one finds
\be
\label{JGRG52}
\frac{V'(\sigma)}{V(\sigma)} \sim - \left(\frac{R_0}{\Lambda_i}\right)^{\frac{2n}{2n+1}}\ ,
\ee
which is very small and  the slow roll condition is satisfied.
Thus, the value of the scalar field $\sigma$ increases very slowly
and the scalar curvature $R$ becomes smaller.
When $\sigma$ becomes large enough and $R$ becomes small enough,
 the inflation could stop.
Another possibility to achieve the exit from the inflation is to add small non-local term
to gravitational action.

We now consider another example, where $f(R)$ satisfies the conditions (\ref{JGRG42}),
(\ref{JGRG44}), and (\ref{JGRG45}) \cite{Nojiri:2007cq}:
\be
\label{JGRG53}
f(R)= \frac{\alpha R^{2n} - \beta R^n}{1 + \gamma R^n} \ .
\ee
Here $\alpha$, $\beta$, and $\gamma$ are positive constants and $n$ is a positive integer.
When the curvature is large ($R\to \infty$), $f(R)$ behaves as
\be
\label{JGRG54}
f(R) \to \frac{\alpha}{\gamma}R^n\ .
\ee
To achieve the exit from the inflation, more terms could be added in the action.
Since the derivative of $f(R)$ is given by
\be
\label{JGRG55}
f'(R) = \frac{n R^{n-1}\left( \alpha \gamma R^{2n} - 2\alpha R^n - \beta \right)}
{\left(1+\gamma R^n \right)^2}\ ,
\ee
we find the curvature $R_0$ in the present universe, which satisfies the condition
$f'(R_0)=0$, is given by
\be
\label{JGRG56}
R_0=\left\{ \frac{1}{\gamma}\left(1+ \sqrt{ 1 + \frac{\beta\gamma}{\alpha} }\right)\right\}^{1/n} \ ,
\ee
and
\be
\label{JGRG57}
f(R_0) \sim -2 \tilde R_0 = \frac{\alpha}{\gamma^2}\left( 1 + \frac{\left(1 - \beta\gamma/\alpha \right)
\sqrt{ 1 + \beta\gamma/\alpha}}{2 + \sqrt{ 1 + \beta\gamma/\alpha}} \right) \ .
\ee
Let us check if we can choose parameters to reproduce realistic
cosmological evolution.
As a working hypothesis, we assume $\beta\gamma/\alpha \gg 1$, then
\be
\label{JGRG58}
R_0 \sim \left(\beta/\alpha\gamma\right)^{1/2n}\ ,\quad
f(R_0)= - 2 \tilde R_0 \sim - \beta/\gamma
\ee
We also assume $f(R_I) \sim (\alpha/\gamma) R_I^n \sim R_I$.
Here $R_I$ is the curvature in the inflationary epoch.
As a result, one obtains
\be
\label{JGRG59}
\alpha \sim 2 \tilde R_0 R_0^{-2n},\ \beta \sim 4 {\tilde R_0}^2 R_0^{-2n} R_I^{n-1},\
\gamma \sim 2 \tilde R_0 R_0^{-2n} R_I^{n-1} .
\ee
Hence, we can confirm the assumption $\beta\gamma/\alpha \gg 1$ if $n>1$
as
\be
\label{JGRG60}
\frac{\beta\gamma}{\alpha} \sim 4 {\tilde R_0}^2 R_0^{-2n} R_I^{2n-2} \sim 10^{228(n - 1)}\gg 1\ .
\ee

Thus, we presented modified gravity models which unify early-time
inflation and late-time acceleration.
One should stress that the above models (\ref{JGRG46}) and (\ref{JGRG53})
satisfy the cosmological
constraints/local tests in the same way as in the HS model \cite{HS}.

\section{Reconstruction of $F(R)$-gravity \label{Sec5}}

In this section, it is shown how we can construct $F(R)$ model realizing
{\it
any} given cosmology
(including inflation, matter-dominated epoch, {\it etc}) using technique
of ref.\cite{NOr}.
The general $F(R)$-gravity action with general matter is given as:
\be
\label{JGRG61}
S = \int d^4 x \sqrt{-g}\left\{F(R) + {\cal L}_{\rm matter}\right\} \ .
\ee
The action (\ref{JGRG61}) can be rewritten by using proper functions $P(\phi)$ and
$Q(\phi)$ of a scalar field $\phi$:
\be
\label{JGRG62}
S=\int d^4 x \sqrt{-g} \left\{P(\phi) R + Q(\phi) + {\cal L}_{\rm matter}\right\}\ .
\ee
Since the scalar field $\phi$ has no kinetic term, one may regard $\phi$
as an auxiliary scalar
field. By the variation over $\phi$, we obtain
\be
\label{JGRG63}
0=P'(\phi)R + Q'(\phi)\ ,
\ee
which could be solved with respect to $\phi$ as $\phi=\phi(R)$. By substituting $\phi=\phi(R)$
into the action (\ref{JGRG62}), we obtain the action of $F(R)$-gravity where
\be
\label{JGRG64}
F(R) = P(\phi(R)) R + Q(\phi(R))\ .
\ee
By the variation of the action (\ref{JGRG62}) with respect to $g_{\mu\nu}$,
 the equation of motion follows:
\be
\label{JGRG65}
0 = -\frac{1}{2}g_{\mu\nu}\left\{P(\phi) R + Q(\phi) \right\}
 - R_{\mu\nu} P(\phi) + \nabla_\mu \nabla_\nu P(\phi)
 - g_{\mu\nu} \nabla^2 P(\phi) + \frac{1}{2}T_{\mu\nu}
\ee
In  FRW universe (\ref{JGRG14}), Eq.(\ref{JGRG65}) has the following form:
\bea
\label{JGRG66}
0&=&-6 H^2 P(\phi) - Q(\phi) - 6H\frac{dP(\phi(t))}{dt} + \rho \nn
0&=&\left(4\dot H + 6H^2\right)P(\phi) + Q(\phi)
+ 2\frac{d^2 P(\phi(t))}{dt} + 4H\frac{d P(\phi(t))}{dt} + p
\eea
By combining the two equations in (\ref{JGRG66}) and deleting $Q(\phi)$, we obtain
\be
\label{JGRG67}
0 = 2\frac{d^2 P(\phi(t))}{dt^2} - 2 H \frac{dP(\phi(t))}{dt} + 4\dot H P(\phi) + p + \rho\ .
\ee
Since one can redefine $\phi$ properly as $\phi=\phi(\varphi)$, we may
choose $\phi$ to be a
time coordinate: $\phi=t$.
Then assuming $\rho$, $p$ could be given by the corresponding sum of
matter with a constant EoS parameters
$w_i$ and writing the scale factor $a(t)$ as $a=a_0\e^{g(t)}$ ($a_0$ : constant),
we obtain the  second rank differential equation:
\be
\label{JGRG68}
0 = 2 \frac{d^2 P(\phi)}{d\phi^2} - 2 g'(\phi) \frac{dP(\phi))}{d\phi} + 4g''(\phi) P(\phi)
+ \sum_i \left(1 + w_i\right) \rho_{i0} a_0^{-3(1+w_i)} \e^{-3(1+w_i)g(\phi)} \ .
\ee
If one can solve Eq.(\ref{JGRG68}), with respect to $P(\phi)$, one can
also find the form of $Q(\phi)$
by using (\ref{JGRG66}) as
\be
\label{JGRG69}
Q(\phi) = -6 \left(g'(\phi)\right)^2 P(\phi) - 6g'(\phi) \frac{dP(\phi)}{d\phi}
+ \sum_i \rho_{i0} a_0^{-3(1+w_i)}  \e^{-3(1+w_i)g(\phi)} \ .
\ee
Thus, it follows that any given cosmology can be realized by some specific
$F(R)$-gravity.

We now consider the cases that (\ref{JGRG68}) can be solved exactly.
A first example is given by
\be
\label{JGRG70}
g'(\phi) = g_0 + \frac{g_1}{\phi}\ .
\ee
For simplicity, we neglect the contribution from matter.
Then Eq.(\ref{JGRG68}) gives
\be
\label{JGRG71}
0 = \frac{d^2 P}{d\phi^2} - \left(g_0 + \frac{g_1}{\phi}\right)\frac{dP}{d\phi}
   - \frac{2g_1}{\phi^2}P \ .
\ee
The solution of (\ref{JGRG71}) is given in terms of the Kummer functions or
confluent hypergeometric functions:
\be
\label{JGRG72}
P=z^\alpha F_K(\alpha,\gamma; z)\ , \quad
z^{1-\gamma} F_K(\alpha - \gamma + 1, 2 - \gamma; z)
\ee
Here
\bea
\label{JGRG73}
&& z\equiv g_0\phi \ ,\quad \alpha\equiv \frac{1+g_1 \pm \sqrt{g_1^2 + 10g_1 + 1}}{4}\ ,\nn
&& \gamma\equiv 1\pm \frac{\sqrt{g_1^2 + 10g_1 + 1}}{2}\ ,\quad
F_K(\alpha,\gamma;z)=\sum_{n=0}^\infty
\frac{\alpha(\alpha + 1)\cdots (\alpha + n -1)}{\gamma(\gamma + 1)\cdots
(\gamma + n - 1)} \frac{z^n}{n!}\ .
\eea
Eq.(\ref{JGRG70}) gives the following Hubble rate:
\be
\label{JGRG74}
H=g_0 + \frac{g_1}{t}\ .
\ee
Then when $t$ is small, $H$ behaves as
\be
\label{JGRG75}
H\sim \frac{g_1}{t}\ ,
\ee
which corresponds to the universe with matter whose EoS parameter is given by
\be
\label{JGRG76}
w=-1 + \frac{2}{3g_1}\ .
\ee
On the other hand, when $t$ is large, we find
\be
\label{JGRG77}
H\to g_0\ ,
\ee
that is, the universe is asymptotically deSitter space.

We now show how we could reconstruct a model unifying the early-time
inflation with late-time acceleration.
In principle, one may consider $g(\phi)$ satisfying the following
conditions:
\begin{itemize}
\item The condition for the inflation ($t=\phi\to 0$):
\be
\label{JGRGa1}
g''(0)=0\ ,
\ee
which shows that $H(0)=g'(0)$ is almost constant, which corresponds to the
asymptotically deSitter
space.
\item The condition fot the late-time acceleration (at $t=\phi\sim t_0$):
\be
\label{JGRGa2}
g''(t_0)=0\ ,
\ee
which corresponds to the asymptotically deSitter
space again.
\end{itemize}
An example could be
\be
\label{JGRGa3}
g'(\phi) = g_0 + g_1 \frac{ \left(t_0^2 - \phi^2 \right)^{n} - t_0^{2n}}
{\left(t_0^2 - \phi^2\right)^{n} + c}\ .
\ee
Here $g_0$, $g_1$, and $c$ are positive constants and $n$ is  positive
integer greater than $1$.
Note that $g'(\phi)$ is a monotonically decreasing function of $\phi$ if
$0<\phi<t_0$
We also assume
\be
\label{JGRGa4}
0<g_0 - \frac{g_1 t_0^{2n}}{c} \ll g_0\ .
\ee
One should note that $g'(0)=g_0$ corresponds to the large Hubble rate in
the  inflationary epoch and
$g'(t_0)=g_0 - \frac{g_1 t_0^{2n}}{c}$ to the small Hubble rate
 in the present universe.
It is very difficult to solve (\ref{JGRG68}) with (\ref{JGRGa3}), so we
expand $g(\phi)$ for
small $\phi$. For simplicity, we consider the case that $n=2$ and no
matter presents. Then
\be
\label{JGRGa5}
g(\phi) = g_0 - \frac{2g_1 t_0^2}{t_0^4 + c} \phi^2 
+ {\cal O}\left(\phi^4\,\mbox{or}\, g_1^2\right)\ .
\ee
Hence, one gets
\bea
\label{JGRGa6}
P(\phi) &=& P_0 + P_1 \e^{g_0\phi} - \frac{2g_1 t_0^2}{t_0^4 + c} \left[ P_1 \left\{ \frac{\phi^3}{3}
 - \frac{3\phi^2}{g_0} + \frac{6\phi}{g_0^2} - \frac{6}{g_0^4} \right\} \e^{g_0\phi}
+ \left\{\frac{2\phi^2}{g_0} + \frac{4\phi}{g_0^2}\right\} P_0 \right. \nn
&& \left. - \frac{P_2}{g_0}\e^{g_0\phi} - P_3 \right] + {\cal O}\left(g_1^2\right)\ .
\eea
Here $P_0$, $P_1$, $P_2$, and $P_3$ are constants of integration.
Using boundary conditions we can specify different modified gravities
which unify the early-time inflation with late-time acceleration.

We may consider another model:
\be
\label{JGRGa7}
g'(\phi) = \frac{a + b \phi^2}{1 + c\phi^2}\ .
\ee
Here $a$, $b$, and $c$ are positive constants satisfying the condition:
\be
\label{JGRGa8}
\frac{b}{c} \ll a\ .
\ee
In the early time $\phi=t\sim 0$, we find
\be
\label{JGRGa9}
H(t)=g'(t)= a + \left(b-ac\right) \phi^2 + \cal{O}\left(t^4\right)\ .
\ee
Then we may identify $a$ as a cosmological constant which generates the inflation.
On the other hand, in the late time $\phi=t\to \infty$, the Hubble rate $H$ is given by
\be
\label{JGRGa10}
H(t) = g'(t) = \frac{b}{c} + \frac{a - b/c}{c\phi^2} + {\cal O}\left(\phi^{-4}\right)\ ,
\ee
which tells that the effective cosmological constant generating accelerating expansion
of the universe could be given by $b/c$.

When $\phi=t$ is small, by comparing (\ref{JGRGa9}) with (\ref{JGRGa5}), we may identify
\be
\label{JGRGa11}
a \leftrightarrow g_0 \ ,\quad ac - b \leftrightarrow \frac{2g_1 t_0^2}{t_0^4 + c} \ .
\ee
Then by using (\ref{JGRGa6}), we find that the corresponding $P$ could be given by
\bea
\label{JGRGa12}
P(\phi) &=& P^I_0 + P^I_1 \e^{a\phi} - \left(ac - b\right) \left[ P^I_1 \left\{ \frac{\phi^3}{3}
 - \frac{3\phi^2}{a} + \frac{6\phi}{a^2} - \frac{6}{a^4} \right\} \e^{a\phi}
+ \left\{\frac{2\phi^2}{a} + \frac{4\phi}{a^2}\right\} P^I_0 \right. \nn
&& \left. - \frac{P^I_2}{g_0}\e^{a\phi} - P^I_3 \right] + {\cal O}\left(\left(ac - b\right)^2 \right)\ .
\eea
Here $P^I_0$, $P^I_1$, $P^I_2$, and $P^I_3$ are constants of integration.
On the other hand, when $\phi=t$ is large, we find
\bea
\label{JGRGa13}
P(\phi) &=& P^L_0 + P^L_1 \e^{b\phi/c} + \frac{a - b/c}{c}\left[
4 P^L_0 \int^\phi d\phi' \e^{B\phi'} \int^{\phi'} \frac{d\phi''}{{\phi''}^3} \e^{-B\phi''} \right. \nn
&& \left. - P^L_1 \left\{ \int^\phi d\phi'
\left( \frac{b}{c\phi'} - \frac{2}{{\phi'}^2} \right)\e^{b\phi'/c}
\right\}\right] + P^L_2 \e^{b\phi/c}\ .
\eea
Here $P^I_0$, $P^I_1$, and $P^I_2$ are constants of integration, again.

The important element of above reconstruction scheme is that it may be
applied partially.
For instance, one can start from the known model which passes local tests
and describes the late-time acceleration. After that, the reconstruction
method may be applied only at very small times (inflationary universe) to
modify such a theory partially. As a result, we get the modified gravity
with necessary early-time behavior and (or) vice-versa.

\section{Dark Matter from $F(R)$-gravity \label{Sec6}}

It is extremely interesting that
dark matter could be explained in the framework of viable $F(R)$-gravity
which was discussed in previous sections.

The previous considerations of $F(R)$-gravity suggest that it may play
the role of gravitational alternative for
dark energy. However, one can study $F(R)$-gravity as a model
for dark matter.
There have been proposed several scenarios to explain dark matter
in the framework of $F(R)$-gravity. In most of such approaches\cite{capo},
the MOND-like scenario or power-law gravity
have been considered. In such scenarios, the field equations
have been solved and the large-scale correction  to the Newton law has
been found  and used as a source of dark matter.

There was, however, an observation \cite{SLAC} that the
 distribution of the matter is different
from that of dark matter in a galaxy cluster. From this it has been
believed
 that the dark matter
can not be explained by the modification of the Newton law
but dark matter should represent some (particles) matter.

It is known that  $F(R)$-gravity contains a particle mode called
`scalaron',
which explicitly appears when one rewrites $F(R)$-gravity in the
the scalar-tensor form (\ref{JGRG23}).
In the Einstein gravity, when we quantize the fluctuations over the
background metric,
we obtain graviton, which is massless tensor particle. In case of
 $F(R)$-gravity, when
one quantizes the fluctuations of the scalar field in the background
metric, one gets the massive scalar
particles in the addition to the graviton. Since the scalar particles in
 $F(R)$-gravity
are massive, the pressure could be negligible and the strength
of the interaction between
such the scalar particles and usual matter should be that of the
gravitational interaction order and therefore
very small. Hence, such scalar particle could be a natural candidate for
dark matter.

In the model \cite{HS} or our models (\ref{JGRG46}) and (\ref{JGRG53}),
 the mass of the effective scalar
field  depends on the curvature or energy density,
in accord with so-called Chameleon mechanism.
As our models (\ref{JGRG46}) and (\ref{JGRG53})  describe the
early-time inflation as well as late-time acceleration, the `scalaron'
particles,
that is, the scalar particles in $F(R)$-gravity,
 could be generated during the inflationary era.
An interesting point is that the mass could change after the inflation
 due to Chameleon mechanism.
Especially in the model (\ref{JGRG46}), the mass decreases
when the scalar curvature increases as shown in (\ref{JGRG49}).
Hence, in the inflationary era, when the curvature is large,
one may consider
the model where $m_\sigma$ is large. After the inflationary epoch, the
scalar particles, generated
by the inflation, could lose their mass.
Since the mass corresponds to the energy, the difference between
 the mass in the
inflationary epoch and that after the inflation could be radiated as
energy and could be converted into
the matter and the radiation.
This indicates that the reheating could be naturally realized in such
model.
Let the mass of $\sigma$ in the inflationary epoch  be $m_I$ and that
after inflation be $m_A$. Then for $N$ particles, the radiated energy $E_N$
may be estimated as
\be
\label{reheat}
E=\left(m_I - m_A\right) N\ ,
\ee
which could be converted into radiation, baryons and anti-baryons (and
leptons).
It is believed that the number of early-time baryons and anti-baryons
 is $10^{10}$ times of
the number of  baryons in the present universe. Since the density of
the dark matter is
almost five times of the density of the baryonic matter, we find
\be
\label{reheat2}
m_I > 10^{10} m_A\ .
\ee

In the solar system, one gets $A=R\sim 10^{-61}\,{\rm eV}^2$. Then if
$n\gg 10\sim 12$ and
$\Lambda_i\sim 10^{20\sim 38}$, the order of the mass $m_\sigma$ is given by
\be
\label{Uf11}
m_\sigma^2 \sim 10^{239\sim 295 - 10n}\,{\rm eV}^2\ ,
\ee
which is large enough so that $\sigma$ could be Cold (non-relativistic) Dark Matter.
On the other hand, in $1/R$-model, the corresponding mass is given by
\be
\label{1overR}
m_{1/R}^2 \sim \frac{\mu^4}{R} \sim 10^{-51}\,{\rm eV}^2\ .
\ee
Here $\mu$ is the parameter with dimension of mass and $\mu \sim 10^{-33}\,{\rm eV}$.
The mass $m_{1/R}$ is very small and cannot be a Cold Dark Matter.
 The corresponding composite
particles can be a Hot (relativistic) Dark Matter but Hot Dark Matter
has been excluded due to
difficulty to generate the universe structure formation.

In the inflationary era, the spacetime is approximated by the de Sitter
space:
\be
\label{dS}
ds^2 = -dt^2 + \e^{2H_0t}\sum_{i=1,2,3} \left(dx^i\right)^2\ .
\ee
 Then the scalar particle $\sigma$
could be Fourier-transformed as
\be
\label{crea1}
\sigma = \int d^3 k \tilde \sigma({\bf k},t) \e^{-i{\bf k}\cdot {\bf x}}\ .
\ee
Hence, the number of the particles with ${\bf k}$ created during inflation
is proportional to
$\e^{\nu\pi}$. Here
\be
\label{crea2}
\nu \equiv \sqrt{\frac{m_\sigma^2}{H_0^2} - \frac{9}{4}}\ .
\ee
Then if
\be
\label{crea3}
\frac{m_\sigma^2}{H_0^2} > \frac{9}{4}\ ,
\ee
sufficient number of the particles could be created.

In the original $f(R)$-frame (\ref{JGRG7}), the scalar field $\sigma$
appears as composite
state. The equation of motion in $f(R)$-gravity contains fourth
derivatives, which means the existence of the extra particle mode or composite state.
In fact, the trace part of the equation of motion (\ref{JGRG13}) has the following
Klein-Gordon equation-like form:
\be
\label{Scalaron}
3\nabla^2 f'(R)=R+2f(R)-Rf'(R)-\kappa^2 T\,.
\ee
The above trace equation can be interpreted as an equation of motion for
the non trivial `scalaron' $f'(R)$. This means that the curvature itself propagates.
In fact the scalar field $\sigma$ in the scalar-tensor form of the theory
can be given by `scalaron',
which is the combination of the scalar curvature in the original frame:
\be
\label{Com}
\sigma = -\ln\left( 1 + f'(R)\right)\ .
\ee
Note that the `scalaron' is different mode from graviton, which is
massless and tensor.

Eq.(\ref{JGRG49}) shows that the mass, which depends on the value of the
scalar field $\sigma$,
 is given by
\be
\label{A5}
m_\sigma^2 \sim \frac{f_0}{f_1^2}\left(\frac{2n+1}{2n+2}\right)
\left(\frac{f_1^2 }{(2n+1)f_0}\right)^{\frac{2n+1}{2n+2}} \sigma^{-\frac{2n+3}{2n+2}}\ .
\ee
If the curvature becomes small, $\sigma$ becomes large and $m_\sigma^2$ decreases.
Then the scalar particles lose their masses after the inflation.
 The difference of the mass in the
inflationary epoch and that after the inflation could be radiated as
energy and can be converted into
the matter and the radiation.

By substituting the expression of $\sigma$  (\ref{JGRG49}) into
(\ref{A5}), one obtains
\be
\label{Inf1}
m_\sigma^2 \sim \frac{f_1^2 A^{2n+3}}{2(2n+1)(n+1)f_0}\ .
\ee
Note that $A$ corresponds to the scalar curvature. Let denote the value
 of $A$ in the inflationary
epoch by $A_I$ and that after the inflation by $A_A$.
 Then the condition (\ref{reheat2}) shows
\be
\label{Inf2}
\frac{m_I}{m_A} \sim \left(\frac{A_I}{A_A}\right)^{n+3/2}> 10^{10}\ .
\ee
For the model with $n=2$, the condition (\ref{reheat2}) or (\ref{Inf2})
could be satisfied
if $A_I/A_A > 10^3$, which seems to indicate that the reheating could be
easily realized
in such a  model.

Now we check if the condition (\ref{crea3}) could be satisfied.
Note $H_0^2 \sim \Lambda_i$.
Eq.(\ref{Inf1}) also indicates that in the  inflationary era, where
$A=R\sim \Lambda_i$,
the magnitude of the mass is given by
\be
\label{Inf3}
m_\sigma^2 \sim \frac{\Lambda_i^{2n+1}}{R_0^{2n}}\ ,
\ee
which is large enough and the condition (\ref{crea3}) is satisfied.
Here Eq.(\ref{JGRG47}) is used.
Thus, sufficient number of $\sigma$-particles could be created.

Let us consider the rotational curve of galaxy.
As we will see the shift of the rotational curve does not occur due to
 correction
to the Newton law between visible matter  (baryon or intersteller gas)
 but due to invisible
(dark) matter, and the Newton law itself is not modified.

Let the temperature of the dark matter be $T=1/k\beta$ where $k$ is the Boltzmann constant.
First, we assume the mass $m_\sigma$ of the scalar particle $\sigma$ is
constant.
As the total mass of  dark matter is much larger than that of baryonic
matter
and radiation, we neglect the contributions from the baryonic matter and radiation just for simplicity.
We now work in Newtonian approximation and the system is spherically symmetric.
Let the gravitational potential, which can be formed by the sum of the dark matter particles,
 be $V(r)$. Then the gravitational force is given by ${\cal F}(r)=-mdV(r)/dr$. If we denote the number
density of the dark matter particles by $n(r)$, in the Newtonian approximation,
by putting $\kappa^2=8\pi G$, one gets
\be
\label{Rot1}
{\cal F}(r)=- \frac{Gm_\sigma^2}{r^2}\int_0^r 4\pi s^2 n(s) ds \,
\ee
and therefore $V(r)$ is given by
\be
\label{Rot2}
V(r)= 4\pi Gm_\sigma \int^r \frac{ds}{s^2} \int_0^s u^2 n(u) du \ .
\ee
If one assumes the number density $n(r)$ of dark matter particles
could obey the Boltzmann
distribution, we find
\be
\label{Rot3}
n(r) = N_0 \e^{-\beta m_\sigma V(r)} \ .
\ee
Here $N_0$ is a constant, which can be determined by the normalization.
Using (\ref{Rot2}) and (\ref{Rot3}) and deleting $n(r)$,
 the  differential equation follows:
\be
\label{Rot4}
\left(r^2 V'(r)\right)'= 4\pi Gm_\sigma N_0 r^2 \e^{-\beta m_\sigma V(r)}\ .
\ee
An exact solution of the above equation is given by
\be
\label{Rot5}
V(r)=\frac{2}{\beta m_\sigma} \ln \left(\frac{r}{r_0}\right)\ ,\quad
r_0^2 \equiv \frac{1}{2\pi Gm_\sigma^2 N_0 \beta}\ .
\ee
As  the general solution for the non-linear differential equation
(\ref{Rot4}) is not known,
we assume $V(r)$ could be given by (\ref{Rot5}). Then the rotational speed $v$ of the stars in the
galaxy could be determined by the balance of the gravitational force
 and the centrifugal force:
\be
\label{Rot6}
m_\star \frac{v^2}{r} = - {\cal F}(r) = m_\star V'(r) = \frac{2m_\star}{\beta m_\sigma r}\ .
\ee
Here $m_\star$ is the mass of a star. Hence,
\be
\label{Rot7}
v^2 = \frac{2}{m_\sigma \beta}\ ,
\ee
that is, $v$ becomes a constant, which could be consistent with the observation.

For the dark matter particles from $f(R)$-gravity, the mass $m_\sigma$ depends on the scalar
curvature or the value of the background $\sigma$ as in (\ref{A5}). The scalar curvature is determined
by the energy density $\rho$ (if pressure could be neglected as in usual baryonic matter and cold dark
matter) and if we neglect the contribution from the baryonic matter,
 the energy density $\rho$ is
given by
\be
\label{Rot8}
\rho(r)=m_\sigma n(r)\ .
\ee
Therefore it follows
\be
\label{Rot9}
m_\sigma = m_\sigma \left(\rho(r)\right) = m_\sigma \left( m_\sigma n(r) \right)\ ,
\ee
which could be solved with respect to $m_\sigma$:
\be
\label{Rot10}
m_\sigma = m_\sigma \left(n(r)\right)\ .
\ee
Furthermore by combining (\ref{Rot3}) and (\ref{Rot10}), one may solve
$m_\sigma$ with respect to
$V(r)$ and $N_0$ as
\be
\label{Rot11}
m_\sigma = m_\sigma \left(N_0,V(r)\right)\ .
\ee
Then (\ref{Rot1}) could be modified as
\be
\label{Rot12}
{\cal F}(r)=- \frac{Gm_\sigma \left(N_0,V(r)\right)}{r^2}\int_0^r
4\pi s^2 m_\sigma \left(N_0,V(r)\right) n(s) ds \,
\ee
which gives, instead of (\ref{Rot4}),
\be
\label{Rot13}
\left(r^2 V'(r)\right)'= 4\pi Gm_\sigma
\left(N_0,V(r)\right) N_0 r^2 \e^{-\beta m_\sigma \left(N_0,V(r)\right) V(r)}\ .
\ee
Eq.(\ref{Rot13}) is rather complicated but at least numerically solvable.

For the model (\ref{JGRG46}), if the curvature is large enough even around the galaxy, the mass $m_\sigma$
is given by (\ref{Inf1}). The scalar curvature $A=R$ is proportional to the energy density (since
the pressure could be neglected), $A\propto \rho$, and the energy
 density $\rho$ is given by
(\ref{Rot8}). Then
\be
\label{Rot14}
n(r) \sim \frac{1}{\kappa^2} \left\{\frac{2(n+1)(2n+1) f_0}{f_1^2}\right\}^{\frac{1}{2n+3}}
\left( m_\sigma (r) \right)^{-\frac{2n+1}{2n+3}}\ .
\ee
Using (\ref{Rot3}), one also gets
\be
\label{Rot15}
V(r) = \frac{2n+1}{(2n+3)\beta m_\sigma(r)}\ln \frac{m_\sigma(r)}{m_0}\ ,\quad
m_0 \equiv \left(\kappa^2 N_0\right)^{-\frac{2n+3}{2n+1}}
\left\{\frac{2(n+1)(2n+1) f_0}{f_1^2}\right\}^{\frac{1}{2n+1}\ .}
\ee
Here $m_0$ has mass dimension.
By substituting (\ref{Rot15}) into (\ref{Rot13}), it follows
\bea
\label{Rot16}
&& \left(\frac{2n+1}{2n+3}\right)\frac{1}{\beta} \left\{
r^2 \left( 1 - \ln \frac{m_\sigma(r)}{m_0} \right)\frac{m_\sigma''(r)}{m_\sigma(r)^2}
 - r^2 \left(3 - 2 \ln \frac{m_\sigma(r)}{m_0} \right)\frac{\left(m_\sigma'(r)\right)^2}{m_\sigma(r)^3}
\right. \nn
&& \left. + 2 r \left( 1 - \ln \frac{m_\sigma(r)}{m_0} \right)\frac{m_\sigma'(r)}{m_\sigma(r)^2}\right\}
= \frac{1}{2} \left\{\frac{2(n+1)(2n+1) f_0}{f_1^2}\right\}^{\frac{1}{2n+3}}
r^2 \left(m_\sigma(r) \right)^{\frac{2}{2n+3}}\ .
\eea
It is very difficult to find the exact solution of (\ref{Rot16}), although
one may solve (\ref{Rot16}) numerically.
Then we now consider the region where $m_\sigma \ll m_0$ but $\ln \left(m_\sigma/m_0 \right)$
is slow varying function of $r$, compared with the power of $r$. In the
region,
we may treat $\ln \left(m_\sigma/m_0 \right)$ as a large negative constant:
\be
\label{Rot22}
\ln \left(m_\sigma/m_0 \right) \sim - C\ .
\ee
Then the following solution is obtained:
\bea
\label{Rot23}
m_\sigma (r) &=& m_0 \left(\frac{r}{r_0}\right)^{- \frac{2(2n+3)}{2n+5}}\ ,\nn
r_0^2 &\equiv & \frac{4(2n+1)(2n + 9)C}{(2n+5)\beta}
\left(\kappa^2 N_0\right)^{\frac{2n+5}{2n+1}}
\left\{\frac{2(n+1)(2n+1) f_0}{f_1^2}\right\}^{-\frac{1}{2n+1}}\ .
\eea
Note that $r_0$ can be real for any positive $n$.
 Eq.(\ref{Rot15}) shows that
\be
\label{Rot20}
V(r) = - \frac{2(2n+1)}{2n+5}\frac{1}{\beta m_0}\left(\frac{r}{r_0}\right)^{\frac{2(2n+3)}{2n+5}}
\ln \frac{r}{r_0}\ .
\ee
Note that the potential (\ref{Rot20}) is obtained by assuming the Newton
law
by summing up the Newton potentials coming from the $f(R)$-dark matter
particles (`scalaron')
distributed around the galaxy.
Eq.(\ref{Rot23}) indicates that the condition $m_\sigma \ll m_0$ requires
$r\gg r_0$.
Then by using the equation for the balance of the gravitational force and the centrifugal force,
as in (\ref{Rot6}), we find
\be
\label{Rot21}
v \propto \left(\frac{r}{r_0}\right)^{\frac{2n+3)}{2n+5}}\ ,
\ee
which is monotonically increasing function of $r$ and the behavior is different from that in (\ref{Rot7}).
If there is only usual baryonic matter without any dark matter,
the velocity is the decreasing
function of $r$, if there is also usual dark matter, as shown in (\ref{Rot7}),
 the velocity is
almost constant, if  dark matter originates from $f(R)$-gravity, as we
consider here, there is
 a region where the velocity could be an {\it increasing} function of $r$.
Of course, one should be more careful as these are qualitative
considerations. The condition $m_\sigma \ll m_0$ requires $r\gg r_0$ but
in the region faraway from galaxy, the scalar curvature becomes small
and the approximation
(\ref{Inf1}) could  be broken. Anyway if there appears a region where
velocity is the increasing
function of $r$, this might be a signal of $f(R)$-gravity origin for
dark matter.
For more precise quantitative arguments, it is necessary to include the
contribution from usual
baryonic matter as well as to do numerical calculation.
In any case, it seems very promising that composite particles from viable
modified gravity which unifies inflation with late-time acceleration may
play the role of dark matter.

\section{Non-minimal modified gravity}

In this section, we consider the theory with non-minimal gravitational coupling as an extension
to the $F(R)$-gravity.

In \cite{NOT}, the non-minimal gravitational coupling of the scalar field
$\phi$
 was considered:
\be
\label{NOT1}
S=\int d^4 x \sqrt{-g}\left[\frac{R}{2\kappa^2} - \frac{f_1(A)}{2}\partial_\mu \phi
\partial^\mu \phi - f_2(A)\right]\ .
\ee
Here $A$ is a geometrical invariant like scalar curvature or Gauss-Bonnet invariant.
In the FRW universe (\ref{JGRG14}), by the variation over $\phi$, one
finds
\be
\label{NOT2}
\dot\phi = q f_1(A)^{-1} a^{-3} \ .
\ee
Here $q$ is a constant. By combining (\ref{NOT2})
 and the gravitational field equation,
 we delete the scalar field $\phi$ and obtain $FRW$-like equation:
\bea
\label{NOT3}
\frac{6}{\kappa^2}H^2 &=& \rho_1 + \rho_2 \ , \nn
\rho_1 &=& \frac{ q^2}{f_1a^6}    \left [
  \frac{1}{2} +3\frac{f_1'}{f_1}(\dot H + 7H^2) +6H \frac{(f_1')^2}{f_1^2} \dot R
  -3H\frac{f_1''}{f_1}\dot R  \right ]\ , \nn
\rho_2 &=& f_2 -6f_2'(\dot H +H^2) +6 H f_2''\dot R\ .
\eea
In case $f_2=0$ and $f_1=f_{HS}$ in (\ref{JGRG8}),
if the curvature $R$ is small,  the
following solution is obtained:
\be
\label{NOT4}
a \propto t^{(k+1)/3}\ ,
\ee
which gives the effective EoS parameter $w_{\rm eff}$ in (\ref{JGRG12}) as
\be
\label{NOT5}
w_{\rm eff}= \frac{1 - n}{1 + n} \ .
\ee
Then the late-time acceleration of the universe $w_{\rm eff}<-1/3$ occurs
 when $n>2$.
Furthermore, the effective phantom regime $w_{\rm eff} <-1$ appears when
$n<-1$.  Within such theory for proper choice of gravitational part (like
in previous sections) one achieves the unification of early-time inflation
with late-time acceleration. Moreover, there is no extra problems to pass
local tests.

One may also consider the theory with non-minimal coupling with
electro-magnetic field \cite{BO}:
\be
\label{BO1}
S=\int d^4 x \sqrt{-g} \left[ \frac{F(R)}{2\kappa^2} - \frac{I(R)}{4}F_{\mu\nu} F^{\mu\nu}
\right] \ .
\ee
Here $I(R)$ is a proper function of the scalar curvature $R$. Even
if  gravity is the
Einstein one ($F(R)=R$), if  $I(R)$ is chosen as
\be
\label{BO2}
I(R) = 1 + f_{HS}(R)\ ,
\ee
($f_{HS}(R)$ is given in (\ref{JGRG8})),  the power law inflation occurs,
where
\be
\label{B03}
a(t) \propto t^{(n+1)/2}\ ,
\ee
when $R/m^2 \gg 1$. Note that the large-scale magnetic fields are
generated
due to the breaking of the conformal invariance of the electromagnetic field through its
non-minimal gravitational coupling.
The main conclusion is that adding the non-minimal coupling with gravity,
the qualitative results of previous sections remain valid.
On the same time, new bounds for non-minimal gravitational couplings only
restrict such couplings to tend to constant (or, sometimes, to zero) at
current epoch. The important point of non-minimal modified gravity is that
as purely gravitational part any $F(R)$ gravity\cite{review,FR,FR1}
may be discussed in cosmological context. The only condition is that it
should be realistic, i.e. to pass the local tests and cosmological bounds.

\section{Discussion \label{Sec7}}

In summary, we reviewed $F(R)$-gravity and demonstrated that some versions
of such theory are viable gravitational candidates for unification of
early-time inflation and late-time cosmic acceleration. It is explicitly
shown that the known critical arguments against such theory do not work
for those models. In other words, the modified gravity under consideration
may pass the local tests (Newton law is respected, the very heavy positive
mass for additional scalar degree of freedom is generated). The
reconstruction of modified $F(R)$ gravity is considered.
It is demonstrated that such theory may be reconstructed for any given
cosmology.
Moreover, the partial reconstruction (at early universe) may be done for
modified gravity which complies with local tests and dark energy bounds.
This leads to some freedom in the choice of modified gravity for the
unification of given inflationary era
compatible with astrophysical bounds and dark energy epoch.
Moreover, non-minimal gravitational coupling with usual matter may be
successfully included into above scheme.
As a final very promising result it is shown that modified gravity under
consideration may qualitatively well describe dark matter, using the
composite scalar particle from $F(R)$ theory and Chameleon scenario.

Thus, modified gravity  remains viable cosmological theory which
is realistic alternative to standard Einstein gravity. Moreover, it
suggests the universal gravitational unification of inflation, cosmic
acceleration
and dark matter without the need to introduce any exotic matter.
Moreover, it remains enough freedom in the formulation of such theory which
is very positive fact, having in mind, coming soon precise observational
data.

\section*{Acknowledgements}

This work is supported in part by the Ministry of Education,
Science, Sports and Culture of Japan under grant no.18549001 and 21st
Century COE Program of Nagoya University provided by the Japan Society
for the Promotion of Science (15COEG01) and in part by MEC (Spain)
projects FIS2006-02842 and PIE2007-50/023 and by RFBR, grant 06-01-00609
(Russia).

\end{document}